\documentstyle[aps,prl,twocolumn]{revtex} %
\begin{document} %
\draft %
\widetext %
%
%
 %
\title{Collisional matter-phase damping in Bose-condensed gas} %
\author{L.\ I.\ Plimak, M.\ J.\ Collett and D.\ F.\ Walls} %
\address{Department of Physics, University of Auckland, \protect \\ %
Private Bag 92019, Auckland, New Zealand} %
\date{\today} %
\maketitle %
\begin{abstract} %
Collisional damping of the excitations in a Bose-condensed  %
gas is investigated over the wide range of  %
energies and temperatures.  %
Numerical results for the damping rate are presented  %
and a number of  %
asymptotic and interpolating expressions for it   %
are derived.  %
\end{abstract} %
\pacs{} %
The experimental realisation of Bose-Einstein condensation  %
in dilute atomic gases \cite{Expts} has led to a new range of  %
experiments investigating the properties of the condensate.  %
Recently,  low energy excitations of a Bose-gas in a  %
magnetic trap have been studied by modulating the trap  %
potential \cite{Ketterle}. The frequencies of the lowest  %
modes agreed well with theoretical predictions based on  %
mean field theory \cite{Stringari}. Higher frequency  %
excitations could be studied, for example, by light scattering  %
\cite{Javanainen}.  %
Very recent experiments  %
\cite{Cornell} have extended the study of low-lying  %
collective excitations to include higher temperatures where a  %
finite non-condensate component may interact with the  %
condensate.  %

A full study of the excitation spectrum,  %
relevant to the recent experimental advances, should include  %
lifetimes of excitations at different temperatures and over  %
the full energy range.  %
In particular the behaviour of the damping rate as  %
a function of temperature needs to be calculated.  %

There have been a number of calculations of the  %
condensate excitations for homogeneous systems,  %
dating back to the sixties and seventies  %
\cite{Beliaev,Popov,Mohling,MaAndWoo,Payne,Hohenberg,Szepfalusi}.  %
These calculations were mainly concerned with  %
understanding the  %
physics of the phase transition.  %
Explicit expressions where found for low-momentum  %
asymptotics of the dumping rate. %
Its    %
high-momentum asymptotics at $T = 0$  %
were calculated by Beliaev \cite{Beliaev}. %

In  %
this paper, we present numerical results for the relaxation of  %
the condensate excitations which encompasses the whole range of  %
energies and temperatures, and derive a number of  %
asymptotic and interpolating expressions for the damping  %
rate. Our results are restricted to the homogeneous  %
case.

We consider Bogoliubov's quasiparticles,  %
described by the field operator $\hat b_{%
{\bf p}%
}$ in the momentum representation,   %
$ %
\hat a_{%
{\bf p}%
} =  %
\left (  %
{\hat b_%
{\bf p}%
  - \alpha _%
{\bf p}%
 \hat b^{\dag}_{-%
{\bf p}%
}} %
\right )/{\sqrt{1 - \alpha _%
{\bf p}%
^2}} . %
$ %
Here, $\hat a_%
{\bf p}%
$ is a particle field operator, and $\alpha _%
{\bf p}%
$  %
is the parameter of Bogoliubov's transformation,   %
$ %
\alpha _%
{\bf p}%
 = 1 + {%
{\bf p}%
^2}/(2 m n_0 U_0) - {\Omega _%
{\bf p}%
 }/(n_0 U_0) %
$, %
where  %
$\Omega _%
{\bf p}%
 = \sqrt{  %
{%
{\bf p}%
^2}/{2m}\left (  %
{%
{\bf p}%
^2}/{2m} + 2 U_0 n_0 %
\right ) %
} %
$ %
is the quasiparticle energy. $U_0$  %
is the parameter of the collisional 
interaction of particles, written in  %
the $S$-wave scattering approximation,   %
$ %
H_{\text{coll}} = ({U_0}/{2})\sum_{%
{\bf p}%
_1 + %
{\bf p}%
_2 = %
{\bf p}%
_3 + %
{\bf p}%
_4} %
\hat a_{%
{\bf p}%
_1}^{\dag}\hat a_{%
{\bf p}%
_2}^{\dag} %
\hat a_{%
{\bf p}%
_3}\hat a_{%
{\bf p}%
_4} %
$, where $U_0 = 4\pi a/m$, $a$ is the scattering length  %
and $m$ is the mass of the particle. We use units where  %
$\hbar = k_{\text{B}} = 1$. %

After the Bogoliubov transformation, 
and neglecting interactions of the  %
quasiparticles not involving the 
condensate, we find the Hamiltonian  %
in the form %
$ %
H = H_0 + H_{\text{int}} %
$, where   %
$ %
H_0 = {\sum _%
{\bf p}%
} \Omega _%
{\bf p}%
 \hat b_{%
{\bf p}%
}^{\dag} %
\hat b_{%
{\bf p}%
} %
$ and   %
\begin{eqnarray} 
\nonumber  
H_{\text{int}}  %
= U_0 \sqrt{n_0} %
{\sum  _{%
{\bf p}%
_1 + %
{\bf p}%
_2 = %
{\bf p}%
_3 }} %
\hat a_{%
{\bf p}%
_1} \hat a_{%
{\bf p}%
_2}\hat a_{%
{\bf p}%
_3}^{\dag}+ \text{H.\ c.}\\  %
\nonumber  
= U_0 \sqrt{n_0} %
{\sum  _{%
{\bf p}%
_1 + %
{\bf p}%
_2 = %
{\bf p}%
_3 }} %
\left [  %
\kappa _n \left (  %
\omega _{%
{\bf p}%
_1}, \omega _{%
{\bf p}%
_2}, \omega _{%
{\bf p}%
_3} %
\right ) %
\hat b_{%
{\bf p}%
_1}\hat  b_{%
{\bf p}%
_2} \hat b_{%
{\bf p}%
_3}^{\dag}  %
\right . \\ \left . + %
\kappa _a \left (  %
\omega _{%
{\bf p}%
_1}, \omega _{%
{\bf p}%
_2}, \omega _{%
{\bf p}%
_3} %
\right ) %
\hat b_{%
{\bf p}%
_1} \hat b_{%
{\bf p}%
_2} \hat b_{-%
{\bf p}%
_3}  %
\right ] %
+ \text{H.\ c.} %
\end{eqnarray}
(in all sums over quasiparticle 
states zero momenta are excluded).  %
For brevity, we omit explicit formulae for  %
the normal and anomalous interaction  %
formfactors $\kappa _n$ and $\kappa _a$ \cite{Full}.  %
They are expressed in terms of  %
the three scaled quasiparticle energies,  %
$\omega _{%
{\bf p}%
_k} = \Omega _{%
{\bf p}%
_k} / \Omega _0,  %
k=1,2,3$, using $ \alpha _%
{\bf p}%
 = \sqrt{ %
1 + \omega _%
{\bf p}%
 ^2} - \omega _%
{\bf p}%
$, where   %
$\Omega _0 = U_0 n_0 = 4\pi a n_0/m$ %
is  %
the energy  %
scale  %
introduced by the  %
Bogoliubov's transformation.  %

To justify the $S$-wave scattering approximation,  %
one needs   %
$a^3 n_0 \ll 1$.  %
Since  $U_0 n_0^{1/2} = 
(4\pi /m)^{3/4}(\Omega _0 a^3 n_0)^{1/4}$, 
this also justifies the limit of non-interacting   %
quasiparticles, when 
$H_{\text{int}} \rightarrow  0$ while $\Omega _0$  %
is fixed,  %
and allows one to  %
consider $H_{\text{int}}$ as a small perturbation. %
Note that, in practice,  %
the  %
density of the condensed phase  %
and  %
the  %
temperature of the sample  %
are functions of  %
experimental conditions \cite{Cornell},  %
rather than the former being  %
a function of the latter.  %
We therefore regard $n_0$ as an independent  %
parameter.  %

The system can be characterised by the normal average %
and the linear response  %
function \cite{Kubo}, 
\begin{mathletters} %
\label{NK}%
\begin{eqnarray} 
\label{NClQ}%
N_%
{\bf p}%
 (t-t') &=&  %
\left \langle  %
\hat b^{\dag}_%
{\bf p}%
 (t')  %
\hat b_%
{\bf p}%
 (t)  %
\right \rangle  %
, %
\\ %
\label{%
RespClQ%
}%
K_%
{\bf p}%
 (t-t')  %
&=& - i \theta (t) \left \langle \left [  %
\hat b_%
{\bf p}%
 (t) , \hat b^{\dag}_%
{\bf p}%
 (t')  %
\right ] \right \rangle .  %
\end{eqnarray}
\end{mathletters}
We neglect anomalous averages (due to the  %
anomalous $bbb$ interaction) because  %
they can be important only at extremely  %
low energies.  %

For non-interacting quasiparticles, %
\begin{eqnarray} 
\label{%
ZeroKN%
}%
K^{(0)}_%
{\bf p}%
 (t) = -i \theta (t)  %
\text{e}^{%
-i{\Omega _%
{\bf p}%
}t%
}%
, \ \  %
N^{(0)}_%
{\bf p}%
 (t) = n_%
{\bf p}%
\text{e}^{-i{\Omega _%
{\bf p}%
}t}, %
\end{eqnarray}
where  %
$ %
n_%
{\bf p}%
 =  %
1/\left (  %
\text{e}^{\Omega _%
{\bf p}%
/T} - 1 \right ) %
$. %
The chemical potential of the 
quasiparticles is zero  %
\cite{Landau} since  %
their number is not an integral of the motion.  %
$K^{(0)}_%
{\bf p}%
 (t)$ is  %
the retarded Green's function of the  %
`free' Schr\"odinger equation,   %
\begin{eqnarray} 
\label{%
DiffK0%
}%
(i\partial / \partial t - \Omega _%
{\bf p}%
)K^{(0)}_%
{\bf p}%
 (t) = \delta (t). %
\end{eqnarray}

In order to derive equations for $K_{\bf p}$ and $N_%
{\bf p}%
$, %
consider the Dyson equation for  %
the two-point quantum averages 
in Perel-Keldish techniques  %
\cite{Keldish},   %
\begin{eqnarray} 
\hat G^{\alpha \alpha '}_%
{\bf p}%
  &=& \hat G^{(0)\alpha \alpha '}_%
{\bf p}%
\nonumber  
\\ %
\label{%
DysonG%
}%
&& + \sum_{\alpha '', \alpha ''' = +,-} %
\varepsilon _{\alpha ''} %
\varepsilon _{\alpha '''} %
\hat G^{(0)\alpha \alpha ''}_%
{\bf p}%
\hat \Sigma ^{\alpha '' \alpha '''}_%
{\bf p}%
\hat G^{\alpha ''' \alpha '}_%
{\bf p}%
  . %
\end{eqnarray}
Here, $\alpha ,\alpha ', 
\alpha '' ,\alpha '' = +,-$ are  %
the C-contour indices \cite{Keldish},    %
$\varepsilon _{\pm} = \mp i$,   %
\begin{eqnarray} 
\label{GbyNK}%
G^{\alpha \alpha '}_%
{\bf p}%
 (t - t') %
=  \left \langle T_c \hat b_%
{\bf p}%
 (t_{\alpha })  %
\hat b^{\dag}_%
{\bf p}%
 (t'_{\alpha '}) \right \rangle  %
,  %
\end{eqnarray}
where  %
$T_c$ is the C-contour 
ordering of the field 
operators
\cite{Keldish}, and   %
$G^{(0)}_%
{\bf p}%
 = G_%
{\bf p}%
 |_{H_{\text{int}} = 0}$. %
$\Sigma ^{\alpha  \alpha '}_%
{\bf p}%
 (t) $  %
is the exact self-energy.  %
For brevity, we omit time integrations,  %
regarding the Green's functions and self-energies  %
as kernels of  integral operators in respect  %
of their time arguments.  %
Since   %
$\left \langle \left [  %
\hat b_%
{\bf p}%
 (t) , \hat b^{\dag}_%
{\bf p}%
 (t')  %
\right ] \right \rangle  = i \left [  %
K_%
{\bf p}%
(t-t') - K^*_%
{\bf p}%
 (t'-t) %
\right ] $, the Green's functions  %
(\ref{GbyNK}) %
can be expressed  %
as linear combinations of $K_%
{\bf p}%
$  %
and $N_%
{\bf p}%
$. %
The Dyson equations (\ref{DysonG})  %
then become,   %
\begin{eqnarray} 
\label{%
DysonK%
}%
\hat K_%
{\bf p}%
 = \hat K^{(0)}_%
{\bf p}%
 &+&   %
\hat K^{(0)}_%
{\bf p}%
\hat \kappa _%
{\bf p}%
\hat K_%
{\bf p}%
 , \\ %
\label{%
DysonN%
}%
\hat N_%
{\bf p}%
 = \hat N^{(0)}_%
{\bf p}%
&+&   %
\hat K^{(0)}_%
{\bf p}%
\hat \kappa _%
{\bf p}%
\hat N_%
{\bf p}%
+ %
\hat K^{(0)}_%
{\bf p}%
\hat \sigma _%
{\bf p}%
\hat K^{\dag }_%
{\bf p}%
+ %
\hat N^{(0)}_%
{\bf p}%
\hat \kappa ^{\dag } _%
{\bf p}%
\hat K^{\dag }_%
{\bf p}%
 ,  %
\end{eqnarray}
where  %
\begin{eqnarray} 
\label{Kappa}%
\kappa _%
{\bf p}%
(t) = - i\left [  %
\Sigma _%
{\bf p}%
 ^{++}(t) -  %
\Sigma _%
{\bf p}%
 ^{+-}(t)  %
\right ] \propto \theta (t),  %
\end{eqnarray}
and $\sigma _%
{\bf p}%
(t)$  %
is also a certain linear combination of the  %
self-energy components.  %

A deeper insight \cite{Full} shows that,  %
ultimately, separation  of the equation for  %
$K_%
{\bf p}%
$ is due to microscopic causality.  %
It is also very important from 
a more practical viewpoint.  %
The initial condition $K_%
{\bf p}%
(0) = -i$ is independent  %
of the interaction. Solving (\ref{DysonK}) hence  %
implies evolving the system during a finite time,  %
and a certain simple approximation  %
(e.g., one-loop  
\cite{Payne}) to the susceptibility  %
$\kappa _%
{\bf p}%
$ may well suffice.  %
Conversely, the knowledge of $N_%
{\bf p}%
$ implies  %
that of the steady-state solution, so that simple  %
approximations to $\sigma _%
{\bf p}%
$ are hopeless  %
(cf the fact that noise sources in kinetic equations  %
cannot be found perturbatively \cite{Walls}).  %
The way around this problem is to find $K_%
{\bf p}%
$ and then  %
recover $N_%
{\bf p}%
$ using Kubo's fluctuation-dissipation theorem,  %
thus making (\ref{DysonN}) redundant. %
In this paper, we confine our attention to $K_%
{\bf p}%
$. %

By making use  of  %
(\ref{DiffK0}),  %
Eq.\ \ref{DysonK}  %
may be re-written  %
in the Markov approximation  %
(discussed below)  %
as,  %
\begin{eqnarray} 
\left ( i{\partial }/{\partial t} - \Omega _%
{\bf p}%
 + i \frac{\gamma _%
{\bf p}%
}{2} %
- \Delta _%
{\bf p}%
 \right )K_%
{\bf p}%
 = \delta(t),  %
\end{eqnarray}
where   %
\begin{eqnarray} 
\Delta _%
{\bf p}%
 - i \frac{\gamma _%
{\bf p}%
}{2} = \lim_{\varepsilon \rightarrow 0} 
\int_0^{\infty}dt  %
\text{e}^{(i \Omega _%
{\bf p}%
 - \varepsilon )t}\kappa_%
{\bf p}%
 (t) . %
\end{eqnarray}
Taking the self-energy in the one-loop  %
approximation \cite{Payne} and %
expressing $G^{(0) \alpha \alpha '}_%
{\bf p}%
$ by  %
$K_%
{\bf p}%
^{(0) }$ and $N_%
{\bf p}%
^{(0) }$,  %
after some algebra  %
we find \cite{Full}  %
$\gamma _%
{\bf p}%
 =  %
\gamma _%
{\bf p}%
^{0} +\gamma _%
{\bf p}%
 ^T =  %
\gamma _%
{\bf p}%
^{0} +\gamma _%
{\bf p}%
 ^{T\prime} +\gamma _%
{\bf p}%
 ^{T\prime\prime}$.  Here, %
\begin{mathletters} %
\label{Gamma}%
\begin{eqnarray} 
\label{%
Gamma0%
}%
\gamma _%
{\bf p}%
 ^0 &=& \frac{\gamma _0 p_0}{2 p}  %
\int _0^{\omega _%
{\bf p}%
} d\omega  \,  %
g
\left ( 
\omega _{\bf p} - \omega ,\omega _{\bf p} 
\right ), 
\\ %
\gamma _%
{\bf p}%
 ^{T \prime} %
&=& \frac{\gamma _0 p_0}{p}  %
\int _0^{\omega _%
{\bf p}%
}  d\omega  \, %
\frac{g
\left ( 
\omega _{\bf p} - \omega ,\omega _{\bf p} 
\right )}{ %
\text{e}^{\omega  / \theta }  %
- 1} ,
\label{%
GammaT1%
}%
\\ %
\gamma _%
{\bf p}%
 ^{T \prime\prime} %
&=& \frac{\gamma _0 p_0}{p}  %
\int _0^{\infty} d\omega   %
\left [  %
\frac{g
\left ( 
\omega ,\omega _{\bf p} 
\right ) 
}{\text{e}^{{\omega }/{\theta }} - 1} -  %
\frac{g
\left ( 
\omega ,\omega _{\bf p} 
\right ) 
}{\text{e}^{{ (\omega _%
{\bf p}%
+ \omega )}/{\theta }} - 1} %
\right ] ,
\label{%
GammaT2%
}%
\end{eqnarray}
\end{mathletters}
where  
\begin{eqnarray} 
g
\left ( 
\omega ,\omega _{\bf p} 
\right ) 
=
\frac{ %
\omega  (\omega _%
{\bf p}%
 + \omega ) \kappa _{n}^2(
\omega , 
\omega _{\bf p}, 
\omega _%
{\bf p}%
 + \omega ) %
}{ %
\sqrt{\left (  %
\omega ^2 + 1 %
\right ) \left [  %
(\omega _%
{\bf p}%
 + \omega )^2 + 1 %
\right ]} %
} , %
\end{eqnarray}%
$p =  \left | %
{\bf p}%
 \right | $,  %
$p_0 = \sqrt{2 m U_0 n_0}  = \sqrt{8 \pi a n_0}$,   %
$\gamma _0 = U_0 p_0^3/2\pi = \Omega _0  %
\sqrt{128 \pi a^3 n_0}$ and  %
$\theta  = T/\Omega _0 %
= 2 T m/p_0^2$.  %

Two types of collision processes contribute  %
to $\gamma _%
{\bf p}%
$. The `probe' %
quasiparticle can collide with a condensate  %
particle, producing two quasiparticles.  %
If the thermal population of the final states  %
is neglected, this is in 
essence a classical collision,  %
responsible for $\gamma ^0_%
{\bf p}%
$.  %
Bosonic stimulation of this process by the thermal  %
population of the final states 
results in $\gamma ^{T\prime}_%
{\bf p}%
$.  %
The `probe' quasiparticle can also collide with  %
another quasiparticle, the final 
state being a condensate  %
particle and a quasi-particle. 
This process is due to the  %
bosonic attraction of the condensate, 
and results in  
$\gamma ^{T\prime\prime}_%
{\bf p}%
$. %
Note that both processes contributing to  %
$\gamma ^{T}_%
{\bf p}%
$ %
are purely  %
quantum. %

The result of direct numerical evaluation  %
of expressions (\ref{Gamma}) is 
shown in Fig.\ \ref{Fig1}.  %
We see that both $\gamma ^{0}_%
{\bf p}%
$ and $\gamma ^{T}_%
{\bf p}%
$ %
vanish if $p \rightarrow 0$, 
hence so does $\gamma _%
{\bf p}%
$.  %
This is important for consistency, 
because (i) the condensate 
should not be damped  %
and (ii) the low-energy excitations 
are physically indistinguishable  %
from it. Note that for low energies 
the thermal contribution always  %
prevails.  %
For higher energies, $\gamma ^{0}_%
{\bf p}%
$ grows monotonically  %
while $\gamma ^{T}_%
{\bf p}%
$ has a maximum at a certain momentum.  %
It is interesting that, as 
the temperature grows, this maximum  %
stabilises at $\Omega _%
{\bf p}%
 \sim \Omega _0$, not at $\Omega _%
{\bf p}%
 \sim T$,  %
as might  be expected.  %
The maximal value of $\gamma ^{T}_%
{\bf p}%
$ is  %
close to $\gamma _0 \theta  =  %
T \sqrt{128 \pi a^3 n_0} \lesssim T$.  %
For energies high enough,  %
the thermal contribution becomes negligible.  %

It is easy to check that  %
$ %
\kappa _{n}^2(\omega _{1},\omega _{2},\omega _{1}+\omega _{2}) %
\approx {9  \omega _{1} \omega _{2}(\omega _{1}+\omega _{2}) }/{32}$ if $ %
\omega _{1}+\omega _{2} \ll 1 $, $ %
\approx y_2$ if $\omega _{1} \gg 1 $ and  $ %
\approx {{\omega _{1}{y_2^2}{{\left( 3 + {y_2^2} \right) }^2}} %
/ {8{{\left( 1 + {y_2^2} \right) }^2}}}$ if $ %
\omega _{1} \ll  \omega _{2}, 1 $,   %
where   %
$y_2 =  %
(\sqrt{\omega _{2}^2 + 1} - 1 )/{\omega _{2}}$. %
We then have,  %
\begin{eqnarray} 
\gamma ^0_%
{\bf p}%
 = \frac{3\gamma _0}{80}\,\left (  %
\frac{p}{p_0} %
\right )^5 %
= \frac{3 p^5}{320 \pi m n_0},  %
\ p \ll p_0, \\ %
\label{Tuned}%
\gamma ^0_%
{\bf p}%
 = \frac{\gamma _0}{2}\,  %
\frac{p}{p_0}\left ( 1 -  %
\delta _%
{\bf p}%
 \right ) %
= 8 \pi a^2 n_0 \frac{p}{m} %
\left ( 1 -  %
\delta _%
{\bf p}%
 \right ),\ p \gg p_0 ,  %
\end{eqnarray}
where  %
$\delta _%
{\bf p}%
 = \frac{2\ln{(p^2/p_0^2)}} %
{p^2/p_0^2}$. %
Except $\delta _%
{\bf p}%
$, these are the well known results 
of Beliaev \cite{Beliaev}.  %
Since the expression for $\delta _%
{\bf p}%
$  %
is valid only for large momenta, 
we are free to `tune' it at low momenta  %
so as to improve its agreement 
with the exact numerical result.  %
On `tuning', $\delta _%
{\bf p}%
 = \frac{2 \ln[p^2/p_0^2+2]}{p^2/p_0^2+2 \ln 2}$.  %
In Fig.\ \ref{Fig2}, we compare 
the results of the direct numerical calculation  %
with the approximate expressions.  %
We see that relation (\ref{Tuned}) 
gives a good approximation to the  %
numerical result for $p \gtrsim p_0$. (It even %
correctly  reproduces  %
the $p^5$ law for low momenta.)  %

Consider now the thermal contribution to  %
the width.  %
For $\omega _%
{\bf p}%
 \ll 1, \theta $, we find  %
$ %
\gamma ^T_%
{\bf p}%
 \approx \gamma _0 f(\theta ) p/p_0 %
$, where  %
\begin{eqnarray} 
\label{Deriv}%
f(\theta ) = \frac{2}{\theta } %
\int _0 ^{\infty} d\omega  \frac{\omega ^2 %
{{{y^2}{{\left( 3 + {y^2} \right) }^2}} %
\text{e}^{\omega / \theta } %
}}{ %
{8 \left (  %
\omega ^2 + 1 %
\right ){{{\left( 1 + {y^2} \right) }^2}}} %
\left (  %
\text{e}^{\omega / \theta } - 1 %
\right )^2},  %
\end{eqnarray}
and    %
$y =   %
(\sqrt{\omega ^2 + 1} - 1 )/{\omega }$.  %
For low temperatures $\theta \ll 1$   %
($T  \ll 8\pi a n_0$),  %
$f(\theta ) = (3\pi ^4/20) \theta ^4 $, and %
\begin{eqnarray} 
\label{%
DerivLowT%
}%
\gamma ^T_%
{\bf p}%
 =  %
\frac{3 \pi ^4  \theta ^4 \gamma _0} %
{20} \,  \frac{p}{p_0}  %
=  %
\frac{3\pi pT^4 m^3}{320 a^2 n_0^3}. %
\end{eqnarray}
$\gamma ^T_%
{\bf p}$ dominates if   %
$\Omega_%
{\bf p}%
 \lesssim T $.  %
With the sound velocity $u = 2 \sqrt{\pi a n_0}/m$, %
(\ref{DerivLowT}) coincides with $\gamma ^T_%
{\bf p}%
 /2 = 3\pi ^3 p T^4/40 n_0 m u^4$,  %
found by Popov \cite{Popov}.  %
For high temperatures $\theta \gg 1$,  %
$f(\theta ) \approx 0.60 \theta $  %
(cf \cite{Payne}) and  %
$\gamma _%
{\bf p}%
^T \sim  %
0.60 \gamma _0 \theta p/p_0$.  %
Note that if $\theta \gg 1$, 
the thermal contribution  %
prevails for low momenta 
$ p \ll p_0$ (cf Fig. \ref{Fig1}). %

A simple semi-quantitative 
expression for $\gamma ^T_%
{\bf p}%
 $,  %
valid if $\theta, \omega _%
{\bf p}%
 \gg 1$, %
can be obtained by dropping %
the factors 
$g
\left ( 
\omega ,\omega _{\bf p} 
\right ) $ 
and 
$g
\left ( 
\omega _{\bf p}- \omega ,\omega _{\bf p} 
\right ) $ 
in  %
the integrands (\ref{GammaT1}), (\ref{GammaT2}),  %
which truncate the  %
integrals  %
in the low-energy region $\omega \lesssim  1$,  %
setting instead  %
the lower limit to some 
$\varepsilon > 0$. The integration is  %
easily performed; 
we then find that we must chose  %
$\varepsilon  = 2$  to 
match the asymptotical behaviour  %
of $\gamma ^T_%
{\bf p}%
$ at  %
$p \rightarrow \infty$. %
Then, 
\begin{eqnarray} 
\frac{\gamma ^T_%
{\bf p}%
}{\gamma ^0_%
{\bf p}%
}  %
\approx \frac{8 m T}{p^2} %
\,\ln \frac{Tm\left (  %
1 - \text{e}^{-p^2/2mT} %
\right )}{8\pi a n_0} . %
\end{eqnarray}
Thus the thermal contribution dominates if  %
$\Omega_%
{\bf p}%
 \lesssim T \ln \left (  %
Tm/8\pi a n_0 %
\right ) \lesssim  T \left |  %
\ln n a^3 \right | $,  %
where $n$ is the total density of particles.   %
The last inequality is due to 
the fact that the temperature  %
should be below the condensation point,  %
$T < T_0 \sim n^{2/3}/m$. %

In Fig.\ \ref{Fig4}, we compare the exact numerical  %
results for the thermal component of the width  %
with the approximate expression for  %
low and high momenta. We have used a `tuned' %
expression to give  %
better accuracy at average momenta,  %
$p \gtrsim p_0$, namely,   %
\begin{eqnarray} 
\label{%
HighEnergyTuned%
}%
\gamma ^T_%
{\bf p}%
 & \approx &  %
2 \gamma _0 \theta \frac{p}{p_0 \omega _%
{\bf p}%
}\ln  %
\frac{ %
1 - \text{e}^{ - (2 + p^2/p_0^2) /\theta } %
}{1 - \text{e}^{- 2/\theta } %
} . %
\end{eqnarray}
This expression coincides, to a very good   %
accuracy, with the numerics for $\omega _%
{\bf p}%
 \geq 1$  %
and $\theta \geq 10$.  %
Together, expressions (\ref{Deriv}) 
and (\ref{HighEnergyTuned})  %
provide a good approximation to $\gamma ^T_%
{\bf p}%
$  %
if the temperature is not too low, $\theta \geq 10$.  %

Consider now the validity of the Markov  %
approximation. Assume that the temperature  %
is not too low, $T \gtrsim \Omega _0$.  %
The bosonic distribution $n_%
{\bf p}%
$ is  %
divergent at low energies. This is  %
overcome by the `truncating' factors  %
$g
\left ( 
\omega ,\omega _{\bf p} 
\right ) $ 
and 
$g
\left ( 
\omega _{\bf p}- \omega ,\omega _{\bf p} 
\right ) $ 
in the integrals, resulting in the  %
major contribution to $\gamma _%
{\bf p}%
 ^T$  %
coming from the energies $\sim \Omega _0$.  %
{\em Ipso facto\/}, it can only be  %
applicable to time scales longer than  %
$1/\Omega _0$. With the maximal value of  %
$\gamma _%
{\bf p}%
 ^T \sim  \gamma _0 \theta $,  %
our results apply if  %
$\theta \ll \Omega _0/\gamma _0$.  %
This can be written as   %
\begin{eqnarray} 
\label{%
Markov%
}%
m T \ll \left (  %
\frac{n_0}{a} %
\right )^{1/2}. %
\end{eqnarray}
To understand what this restriction means  %
in terms  %
of the critical temperature $T_{c}$, note  that %
$m T \sim n^{\prime 2/3}$, where $n'$ is  %
the density of the non-condensate phase.  %
Estimate (\ref{Markov}) then is equivalent  %
to  %
$n' \ll n_0/(n_0 a^3)^{1/4}$. This does  %
not contradict $n' \sim n_0$.  %
Thus, although certainly $T < T_{c}$, a  %
stronger condition $T \ll  T_{c}$ %
does not seem necessary.  %

The authors are grateful to 
Prof.\ R.\ Graham for valuable discussions.  %
This work was supported by the Marsden Fund and the 
University of Auckland Research Fund. 

 %
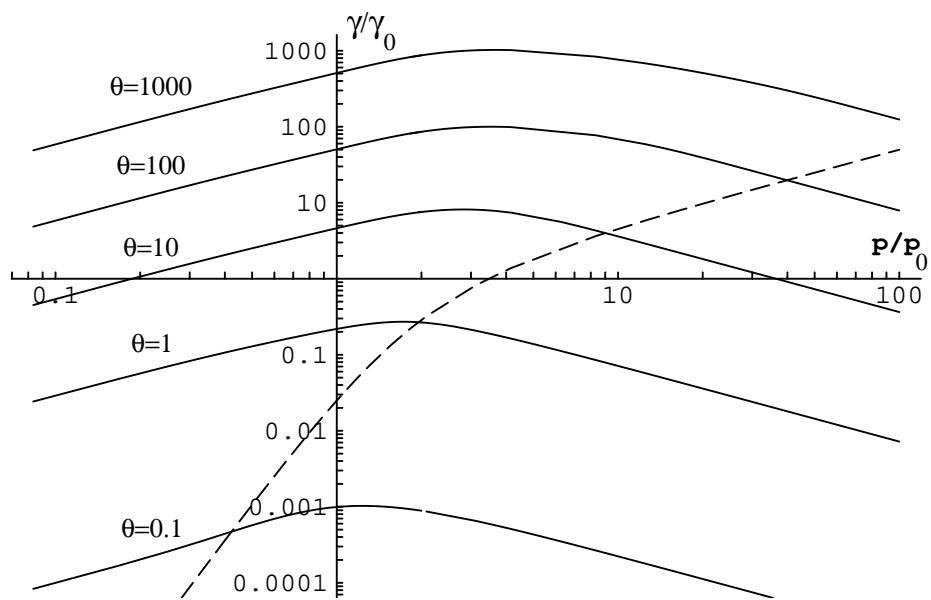
\begin{figure} %
\caption{Results of the direct numerical calculation of  %
the width components $\gamma ^0_%
{\bf p}%
$ (dashed line) and $\gamma ^T_%
{\bf p}%
$ (solid lines).} %
\label{Fig1} %
\end{figure} %
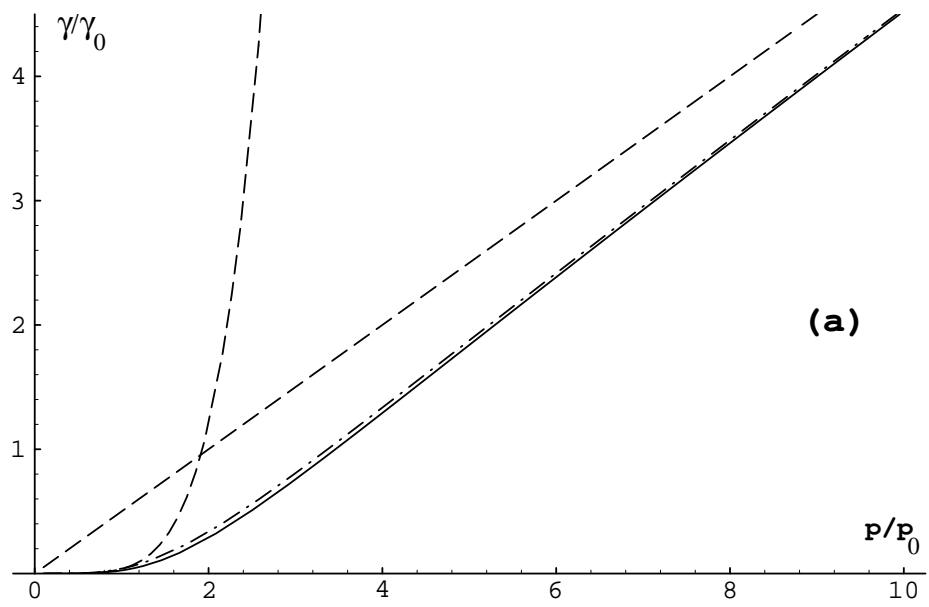
\begin{figure} %
\caption{Comparison of the results of the direct numerical calculation  %
of $\gamma ^0_%
{\bf p}%
$ (solid line) %
with the approximate expressions:  %
dashed lines -- Beliaev's asymptotics for small and large momenta,  %
dash-dotted line -- `tuned' expression (\protect \ref{Tuned}).} %
\label{Fig2} %
\end{figure} %
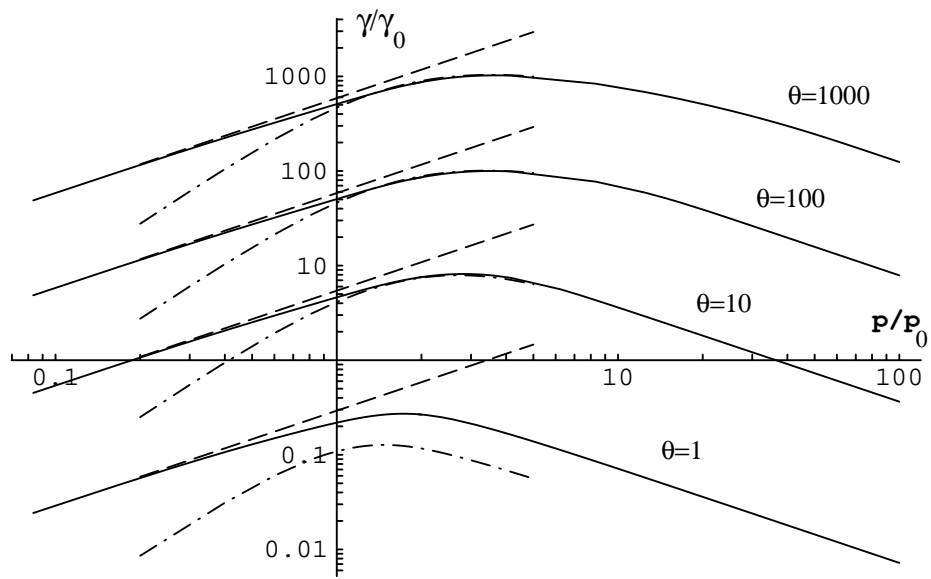
\begin{figure} %
\caption{Comparison of the results of the direct numerical calculation  %
of $\gamma ^T_%
{\bf p}%
$ (solid lines) %
with approximate expressions  %
(\protect \ref{Deriv}) (dashed lines) %
and (\protect \ref{HighEnergyTuned}) (dash-dotted lines).} %
\label{Fig4} %
\end{figure} %
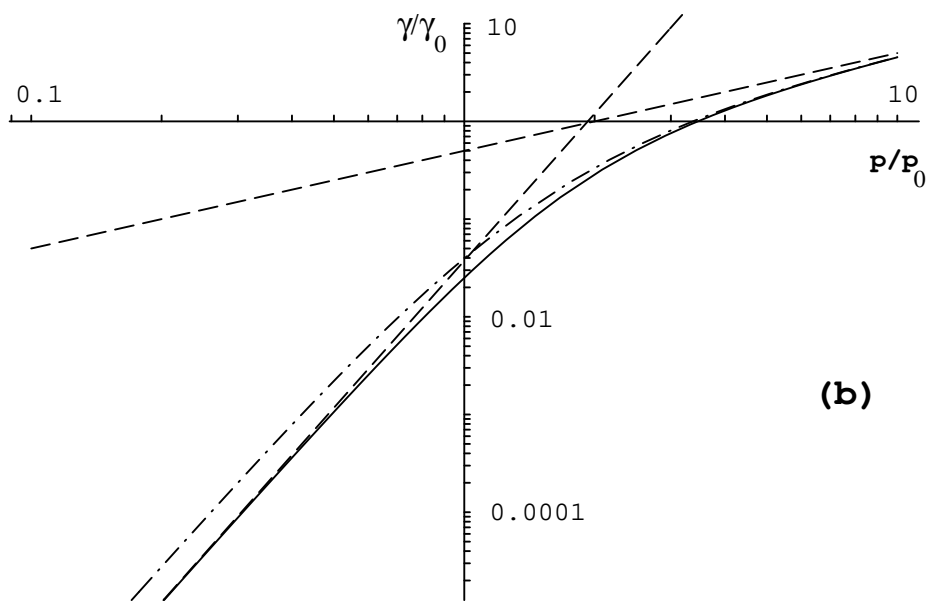
\end{document}